# Realization of $5\frac{h}{e^2}$ with graphene quantum Hall resistance array

Jaesung Park[a)], Wan-Seop Kim[b)], and Dong-Hun Chae[c)]

*Korea Research Institute of Standards and Science, Daejeon 34113, Republic of Korea*

[a)]Electronic mail: jspark99@kriss.re.kr

[b)]Electronic mail: ws2kim@kriss.re.kr

[c)]Electronic mail: dhchae@kriss.re.kr

**Abstract**

We report on realization of 10 quantum Hall devices in series fabricated using epitaxial graphene on silicon carbide. Precision measurements with a resistance bridge indicates that the quantized Hall resistance across an array at filling factor 2 is equivalent to $5\frac{h}{e^2}$ within the measurement uncertainty of approximately $4\times10^{-8}$. A quantum-Hall phase diagram for the array shows that a metrological quantization of $5\frac{h}{e^2}$ can be achieved at the magnetic field of 6 T and temperature of 4 K. This experiment demonstrates the possibility of timely unchangeable resistance reference in various ranges in relaxed experimental conditions.

The integration of quantum devices plays a key role in electrical metrology. The Josephson voltage standard is a prime example. A single Josephson junction can generate only a few millivolts. State-of-the-art integrated chips with tens of thousands Josephson junctions, by contrast, span the voltage range up to $\pm 10$ V[1,2]. The enlargement of voltage is important for calibration because commercial instruments, including digital voltmeters and the Zener voltage references, commonly have the range up to 10 V. Also, an array of quantum Hall devices in series or parallel can extend the resistance range from 1 Ω to 10 MΩ [3, 4, and references therein] from the fixed quantum of resistance ($\frac{h}{e^2}$), even though an intrinsic nonquantum mechanical interconnection resistance exists, which can be minimized below the measurement uncertainty by the triple connection technique[5]. Unlike the Josephson voltage array operating at the liquid helium temperature of 4.2 K, the realization of quantum resistance based on GaAs/AlGaAs heterostructures is demanding. The required magnetic field and temperature are typically approximately 10 T and below 1.5 K, respectively. Therefore, integrated quantum Hall resistors may not be used as widely as the Josephson voltage array in reality.

Graphene was discovered to be a promising material for quantum Hall applications. The emergent quantum Hall effect of graphene in relaxed experimental conditions stems from its distinct electronic energy band. Its linear energy-momentum dispersion results in a characteristically different Landau quantization of massless Dirac fermions from the quadratic dispersion in semiconductors [6, 7]. Consequently, the order-of-magnitude larger energy spacing between Landau levels allows the quantum Hall state at a lower magnetic field ($\leq$ 3.5 T) and higher temperature (> 4 K) for metrological purposes[8]. In practice, however, even though the uniform graphene growth and doping control have been demonstrated for the metrological requirements [8-10], the long-term stability at the ambient conditions remains to be overcome for the metrological applications.

Integration of graphene quantum Hall devices has been demonstrated with a manual interconnection [11] or with the standard microfabrication technique [12]. The relative deviations from the expected calculable nominal values, however, became up to $1\times10^{-4}$ [12] even though a graphene array is expected to be operated more efficiently. It was discovered that the imperfect growth [12] at a large scale prohibits the ideal realization of quantum resistance via integration as well as the energy dissipation in the reduced Hall devices. Another challenges are the high-yield interconnection and minimized leakage current between cross-over interconnections in an integrated circuit. Recently, the interconnection resistance has been minimized by employing a superconducting interconnection [13] with the zero resistance and a split-contact geometry [14] between graphene and metal lead with a reduction of contact resistance by orders of magnitude.

Here, we report on realization of an array of 10 quantum Hall devices in series fabricated using epitaxial graphene on silicon carbide (SiC). Unlike previous reports [11, 12] on the graphene quantum Hall resistance (QHR) array, a homogeneous growth of large-area epitaxial graphene (EG) and a reliable interconnection result in a quantized Hall array resistance (QHAR) close to the expected nominal value. Precision measurements using a bridge technique confirm that the achieved value of QHAR at filling factor 2 agrees with $5\frac{h}{e^2}$ within a relative measurement uncertainty of approximately $4\times10^{-8}$. For metrological applications, we investigated a quantum-Hall phase diagram with respect to the magnetic field and temperature. The quantization of $5\frac{h}{e^2}$ can be realized at the magnetic field of 6 T and temperature above 4 K. This simple demonstration shows that graphene-based QHAR is applicable to metrology with favorable experimental conditions and more component Hall devices with an employed dimension can be integrated for various resistance values.

Epitaxial graphene was grown on the silicon (Si) face of semi-insulating 4H-SiC by sublimating Si atoms in an argon atmosphere of 750 torr at 1600 °C for approximately 5 minutes [15]. To prevent the step bunching on SiC surface with a slow Si-sublimation, we employed a modified graphite susceptor with a small gap as well as the polymer-assisted sublimation method [16]. It is well known that the reduction in SiC terrace height is crucial to achieve the quantum Hall state of EG without dissipation [8,9, 16-18]. Figure 1(a) shows the morphology of EG measured using atomic force microscopy (AFM) in the noncontact mode. Smooth EG with a height of terrace below 1 nm was observed. This indicates that the growth condition is suitable for controlling the terrace structure. Another decisive factor to determine the quality of EG is the monolayer coverage because a bilayer region becomes insulating at a low carrier density and may deteriorate the quantum Hall edge state [19]. Particularly, the growth of large-area monolayer graphene without bi- or multilayer regions is essential to produce a reliable graphene QHR array comprising many Hall components.

Raman spectroscopy is a well-known method to distinguish the number of layers in a graphitic structure [20]. The shape and a full-width at half-maximum (FWHM) of 2D-band of the Raman spectrum provide information regarding the number of layers in EG [21]. Therefore, a map of 2D-FWHM acquired by confocal scanning Raman spectroscopy at a large scale is typically employed to prove the uniformity of EG on SiC. We used a 532-nm-laser source with a power of approximately 20 mW. The upper inset of Fig. 1 (b) exhibits a color map of the FWHM of the 2D-band. The lower inset of Fig. 1(b) shows the Raman spectra of 2D-bands of EG on three randomly selected positions, and every 2D-band is fitted by a single Lorentzian function with an FWHM of approximately 35-36 cm$^{-1}$. Fig. 1(b) shows a histogram of the 2D-FWHM. The uniform green-colored map of 2D-FWHM and the single Gaussian fitting (red curve) with a center of 35.4 cm$^{-1}$ and width of 4.9 cm$^{-1}$ to the histogram of the 2D-FWHM are direct evidences of the homogeneous growth of EG. To ensure the uniformity in the 7 mm × 7 mm area of EG grown on SiC, we performed Raman mapping on different positions near four edges and obtained the similar maps of 2D-FWHM (see the Raman maps in Fig. S1 in the supplementary material.).

To realize a high-value quantum resistance of $5\frac{h}{e^2}$, we designed a graphene QHR array comprising 10 single Hall devices connected in series using the triple connection technique [5], as depicted in Fig. 1(c). The triple connection technique, which can minimize the interconnection resistance due to an interconnecting metal line and a contact between two-dimensional electron system and lead, has been already proven useful in GaAs-based QHR arrays [3, 22, 23]. The geometry (width and thickness) of the interconnecting metal line and the target contact resistance are determined as follows. For instance, the interconnection resistance of 25 Ω (a contact resistance of 20 Ω and a designed metal-line resistance of 5 Ω) results in the relative deviation ($\frac{R_{QHAR}-10R_H}{10R_H}$) of 0.01 μΩ/Ω from the nominal value of $5\frac{h}{e^2}$, calculated by the following simplified equation describing 10-QHRs in series with the triple connection geometry[22].

$$\frac{R_{QHAR}-10R_H}{10R_H} \approx \frac{10-1}{4\times 10}\left(\frac{R_{interconnet}}{R_H}\right)^3.$$

Here, $R_H$ and $R_{interconnet}$ are the quantum Hall resistance ($\frac{h}{2e^2}$) at filling factor 2 and the interconnection resistance comprising two contact resistances and a resistance of an interconnecting metal line, respectively. See the resistance characterization of employed metal line geometry in Fig. S2 in the supplementary material.

Electron-beam lithography was employed for the device fabrication, including 1)

graphene channeling (width: 50 μm , length: 220 μm ), 2) Pd/Au contacts and Ti/Au interconnecting metal lines, and 3) SiO$_2$ insulation to disconnect the cross-over metal lines. We note that a fabrication of graphene quantum Hall array requires uncommon processes for a low resistance of metal lines and a reliable insulation of cross-over interconnections, in contrast to a single graphene quantum Hall device or GaAs-based counterpart. We described the fabrication details in Fig. S3 and S4 in the supplementary material.

After removing the organic residues on the EG surface during the device fabrication by vacuum annealing at 500 °C, we applied a polymer-assisted hole-doping [24, 25] on EG to reduce the electron density to fulfill the filling factor requirement such that the quantum Hall effect can occur at a lower magnetic field. To perform magneto-transport measurements and precision Hall resistance measurements, the array was loaded in a homemade probe for a cold-finger-type commercial cryostat (Physical Property Measurement System, Quantum Design). The homemade probe is electrically insulated by a sapphire-spacer and custom wires from the electrically noisy commercial cryostat; however, it is still thermally anchored to the cold-finger. Figure 1(d) depicts a quantized Hall resistance ($R_{xy}$) close to $5\frac{h}{e^2}$ (129 064.037 296 522 $\Omega_{2019}$) with a red trace and the corresponding suppression of longitudinal resistance ($R_{xx}$) of a single Hall device with a blue trace above the magnetic field of 5 T. The average electron density (4.6 × 10$^{11}$ cm$^{-2}$) and mobility (6000 cm$^2$ V$^{-1}$s$^{-1}$) of a single quantum Hall device are estimated with the quantum Hall plateau at filling factor 6 at the magnetic field of 1.6 T and resistance of 2.2 kΩ at zero magnetic field.

The combined Hall resistance of the graphene array and the longitudinal resistance of a single Hall device in the array were measured using a cryogenic current comparator resistance bridge based on the Meissner effect. Graphene QHAR was compared with a 10 kΩ resistance reference, which was pre-calibrated with the bridge in comparison to a GaAs quantum Hall resistance standard. To determine the resistance ratio, two currents were fed into two separate coils connected to each resistor as depicted in Fig. 2(a). The turn ratio of the two coils and the direction of the currents were configured to minimize the residual magnetic flux monitored using a superconducting quantum interference device (SQUID) and the bridge voltage difference (Δ$U$) measured using a nanovoltmeter. One of the current sources($I_2$) was adjusted with feedback from the SQUID. The $N_1$ and $N_2$ were 2620 and 203, respectively. For a fine balancing, an auxiliary current, corresponding to a fractional effective turn ($kN_A$), was applied. The currents were driven in bipolar to avoid an off-set voltage and its time drift. Consequently, the resistance ratio was determined by the following equation:

$$\frac{R_1}{R_2} = \frac{N_1 + kN_A}{N_2}\left(1 + \frac{\Delta U}{\Delta(IR)}\right) .$$

Here, $\Delta(IR)$ is the voltage drop across a resistor, typically approximately 1 V. For the longitudinal resistance measurement, a Hall probe in a single Hall device was re-connected to an adjacent probe, as illustrated by the dotted-line in Fig. 2(a). The longitudinal resistance was extracted by subtracting the combined Hall resistance from the Hall resistance obtained in the reconfigured geometry for a given setting of magnetic field and temperature.

The precision of the resistance measurements is limited by the electric insulation of the homemade probe, flux resolution of the employed SQUID, and statistical type-A uncertainty. The overall relative measurement uncertainty becomes approximately $4 \times 10^{-8}$. An insulation of approximately 5 TΩ results in a relative error of approximately 30 parts per billion (ppb). The insulation was determined by a commercial tera-ohmmeter (TO-3 by Fischer Elektronik GmbH). The flux via the $N_1$-turn coil induced by $I_1$ is approximately $1850\phi_o$ from the flux linkage of 11 µA·turns/$\phi_o$. Here, $\phi_o$ is the quantum of flux ($h/2e$). Assuming the flux resolution of the SQUID as 10 µ$\phi_o$ conservatively, the relative measurement error becomes approximately 5 ppb. The statistical type-A uncertainty was typically approximately several ppb. Minor uncertainty contributions include the turn ratio error, voltage measurement error, and auxiliary current source error, which are smaller than 1 ppb. Considering the probability distribution of each error parameter, the extended measurement uncertainty becomes approximately 40 ppb at the 95 % confidence level. The corresponding resistance uncertainty is close to 5 mΩ for the longitudinal resistance. Figure 2(b) depicts that the Allan deviation of the bridge voltage difference, acquired for 10 hours at a magnetic field of 10 T at the base temperature of 2 K, follows the inverse square root time dependence. This indicates that the uncorrelated white noise is predominant in the statistical measurement.

We performed precision measurements of the combined Hall resistance and the longitudinal resistance of a single component Hall device with respect to the magnetic field and temperature to investigate the quantum-Hall phase diagram for metrological criteria. The phase diagram in Fig. 3(a) summarizes the results. Within the region in blue, a deviation of QHAR from the nominal value of $5\frac{h}{e^2}$ is comparable with the measurement uncertainty of $4\times10^{-8}$. We note that the observed deviation may arise from the limited insulation of employed homemade probe, not from the QHR array itself. Outside this region (depicted in red), the deviation increases significantly, relative to the uncertainty, owing to the break-down of the quantum Hall state. In addition, the longitudinal resistance increases significantly. Figures 3(b) and 3(c) show the data-sets acquired with the magnetic field and the temperature fixed at 6 T and 8 K, respectively. The left vertical axis and hexagonal symbol (right vertical axis and circular symbol) represent the relative deviation with respect to $5\frac{h}{e^2}$ (the longitudinal resistance). This result reflects that the precise quantization in the array can be achieved with the magnetic field of 6 T at the liquid helium temperature (4.2 K), which is a key conclusion of this work. This operational combination is similar to that of a single Hall device. It implies that the quality of component Hall devices is reasonably homogeneous, and that the additional device-fabrication procedures for the array do not notably degrade the channeling graphene.

Several issues, must be further investigated for metrological applications, including the current dependence of the quantized Hall array resistance and the long-term stability. The applied current through the graphene QHR is expected to be larger than the employed current of approximately 4 µA by two orders of magnitude[8]. For the current dependence measurement, an applied current to the temperature-controlled 10 kΩ resistance reference in an air bath must be increased accordingly by two orders of magnitude, which may change the value of the resistance reference by Joule heating. The upper limit of the current will be investigated with a resistance reference in an oil-bath temperature regulation in the near future. We mark that the quantization of the Hall resistance does not change up to 12 µA in the present temperature regulation. The long-term stability is an important requirement to be investigated. The polymer-encapsulated molecular-doped array device will be measured in time with array devices stored in an inert gas atmosphere. A high-yield realization of interconnection in an array is another requirement. The present fabrication includes a high temperature annealing, which can yield a higher contact resistance. For instance, if only one contact resistance increases up to 150 Ω with the remaining 59 contacts having a typical resistance of a few Ohms, the relative deviation can reach to 4 parts in $10^8$, which is comparable to the observed value. We will investigate the contact failure by the annealing process statistically.

Quantum mechanical resistor in the various range could be an essential element in quantum electronics like an artifact resistor in the conventional electronics. The invariant resistors realized by Hall devices connected in series/parallel can be conceived for the electronics applications. For instance, a current-to-voltage converter can be realized with a high-value resistance array to measure a small current. Unlike a single GaAs Hall device operational typically at 50 µA, a graphene Hall device can be operated at up to 0.5 mA with the metrological accuracy maintained. Therefore, graphene Hall devices in parallel could allow an unprecedentedly large current for an invariant current source, which can be used in, for instance, a Kibble balance [26] for the realization of the newly defined kilogram. Also, the efficient operation of graphene-based arrays, stemming from its characteristic linear energy-momentum dispersion, would allow more practical applications.

In summary, we experimentally realized a precise quantization of $5\frac{h}{e^2}$ using 10 graphene quantum Hall devices connected in series by the triple connection geometry. Using a resistance bridge, we confirmed that the quantized Hall resistance across the array is equivalent to the expected nominal value within the measurement uncertainty of 4 parts in $10^8$. Furthermore, quantization can be achieved at the magnetic field down to 6 T and temperature of 4 K, which is important for practical applications in the future.

This research was supported by the Research on Redefinition of SI Base Units (KRISS-2020-GP2020-0001) funded by the Korea Research Institute of Standards and Science. This work was supported partially by the Graphene Impedance Quantum Standard (Grant No. NRF-2019K1A3A1A78077479) funded by the National Research Foundation of Korea. This work was supported in part by the Joint Research Project GIQS (18SIB07). This project received funding from the European Metrology Programme for Innovation and Research (EMPIR) co-financed by the Participating States and from the European Unions' Horizon 2020 research and innovation programme.

See the supplementary material for the Raman maps of the uniform growth of epitaxial graphene, the resistance characterization of interconnecting metal-line, the fabrication process, and the optical microscopic image of cross-over interconnection.

**Reference**

[1] J. Niemeyer, J. H. Hinken, and R. L. Kautz, Appl. Phys. Lett. **45,** 478 (1984).

[2] F. L. Lloyd, C. A. Hamilton , J. A. Beall, D. Go, R. H. Ono, and R. E. Harris, IEEE Electron Device Lett. **8**, 449 (1987).

[3] F. P. M. Piquemal, J. Blanchet, G. Genevès, and J.-P. André, IEEE Trans. Instrum. Meas. **48**, 296(1999).

[4] T. Oe, S. Gorwadkar, T. Itatani, and N.-H. Kaneko, Conf. on Precision Electromagnetic Measurements (Paris, France, 8–13 July 2018), 8501182 (2018).

[5] F. Delahaye, J. Appl. Phys. **73**, 7914 (1993).

[6] K. Novoselov, A. K. Geim, S. Morozov, D. Jiang, M. Katsnelson, I. Grigorieva, S. Dubonos, and A. Firsov, Nature **438**, 197 (2005).

[7] Y. Zhang, Y.-W. Tan, H. L. Stormer, and P. Kim, Nature **438**, 201 (2005).

[8] R. Ribeiro-Palau, F. Lafont, J. Brun-Picard, D. Kazazis, A. Michon, F. Cheynis, O. Couturaud, C. Consejo, B. Jouault, W. Poirier, and F. Schopfer, Nat. Nanotech. **10**, 965 (2015).

[9] A. Tzalenchuk, S. Lara-Avila, A. Kalaboukhov, S. Paolillo, M. Syväjärvi, R. Yakimova, O. Kazakova, T. J. B. M. Janssen, V. Fal'ko, and S. Kubatkin, Nat. Nanotech. **5**, 186 (2010).

[10] T. J. B. M. Janssen, A. Tzalenchuk, S. Lara-Avila, S. Kubatkin, and V. I. Fal'ko, Rep. Prog. Phys. **76**,104501 (2013).

[11] S. Novikov, N. Lebedeva, J. Hämäläinen, I. Iisakka, P. Immonen, A. J. Manninen, and A. Satrapinski, J. Appl. Phys. **119**, 174504 (2016).

[12] A. Lartsev, S. Lara-Avila, A. Danilov, S. Kubatkin, A. Tzalenchuk, and R. Yakimova, J. Appl. Phys. **118**, 044506 (2015).

[13] M. Kruskopf, A. F. Rigosi, A. R. Panna, M. Marzano, D. Patel, H. Jin, D. B. Newell, and R. E. Elmquist, Metrologia, **55**, R27(2019).

[14] M. Kruskopf, A.F. Rigosi, A.R. Panna, D.K. Patel, H. Jin, M. Marzano, M. Berilla, D. B. Newell, and R.E. Elmquist, IEEE Trans. Instrum. Meas. **66**, 3973(2019).

[15] K.V. Emtsev, A. Bostwick, K. Horn, J. Jobst, G. L. Kellogg, L. Ley, J. L. McChesney, T. Ohta, S.A. Reshanov, J. Röhrl, E. Rotenberg, A. K. Schmid, D. Waldmann, H. B. Weber, and Seyller, Th. Nat. Mater. **8**, 203 (2009).

[16] M. Kruskopf, D. M. Pakdehi, K. Pierz, S. Wundrack, R. Stosch, T. Dziomba, M. Götz, J. Baringhaus, J. Aprojanz, C. Tegenkamp, J. Lidzba, T. Seyller, F. Hohls, F. J. Ahlers, and H. W. Schumacher, 2D Materials, **3**, 041002 (2016).

[17] T. Schumann, K.-J. Friedland, M. H. Oliveira Jr., A. Tahraoui, J. M. J. Lopes, and H.

Riechert, Phys. Rev. B. **85**, 235402 (2012).

[18] Y. Yang, G. Cheng, P. Mende, I. G. Calizo, R. M. Feenstra, C. Chuang, C.-W. Liu, C.-I. Liu, G. R. Jones, A. R. Hight Walker, and R. E. Elmquist, Carbon **115**, 229 (2017).

[19] C. Chua, M. Connolly, A. Lartsev, T. Yager, S. Lara-Avila, S. Kubatkin, S. Kopylov, V. Fal'ko, R. Yakimova, R. Pearce, T. J. B. M. Janssen, A. Tzalenchuk, and C. G. Smith, Nano. Lett. 14, 3369 (2014).

[20] A. C. Ferrari, J. C. Meyer, V. Scardaci, C. Casiraghi, M. Lazzeri, F. Mauri, S. Piscanec, D. Jiang, K. S. Novoselov, S. Roth, and A. K. Geim, Phys. Rev. Lett. **97**, 187401 (2006).

[21] D. S. Lee, C. Riedl, B. Krauss, K. v. Klitzing, U. Starke, and J. H. Smet, Nano. Lett. **8**, 4320 (2008).

[22] T. Oe, Member, S. Gorwadkar, T. Itatani, and N.-H. Kaneko, IEEE Trans. Instrum. Meas. **66**, 1475(2017).

[23] D.-H. Chae, W.-S. Kim, T. Oe, and N.-H. Kaneko, Metrologia **55**, 645 (2018).

[24] H. He, K. H. Kim, A. Danilov, D. Montemurro, L. Yu, Y.W. Park, F. Lombardi, T. Bauch, K. Moth-Poulsen, T. Iakimov, R. Yakimova, P. Malmberg, C. Müller, S. Kubatkin, and S. Lara-Avila, Nat. Comm. **9**, 3956 (2018).

[25] H. He, S. Lara-Avila, K. H. Kim, N. Fletcher, S. Rozhko, T. Bergsten, G. Eklund, K. Cedergren, R. Yakimova, Y. W. Park, A. Tzalenchuk, and S. Kubatkin, Metrologia **56**, 045004 (2019).

[26] B. P. Kibble, Atomic Masses and Fundamental Constants **5**, ed. J. H. Sanders and A. H. Wapstra (New York: Plenum), 541–51(1976).

**Figure 1**

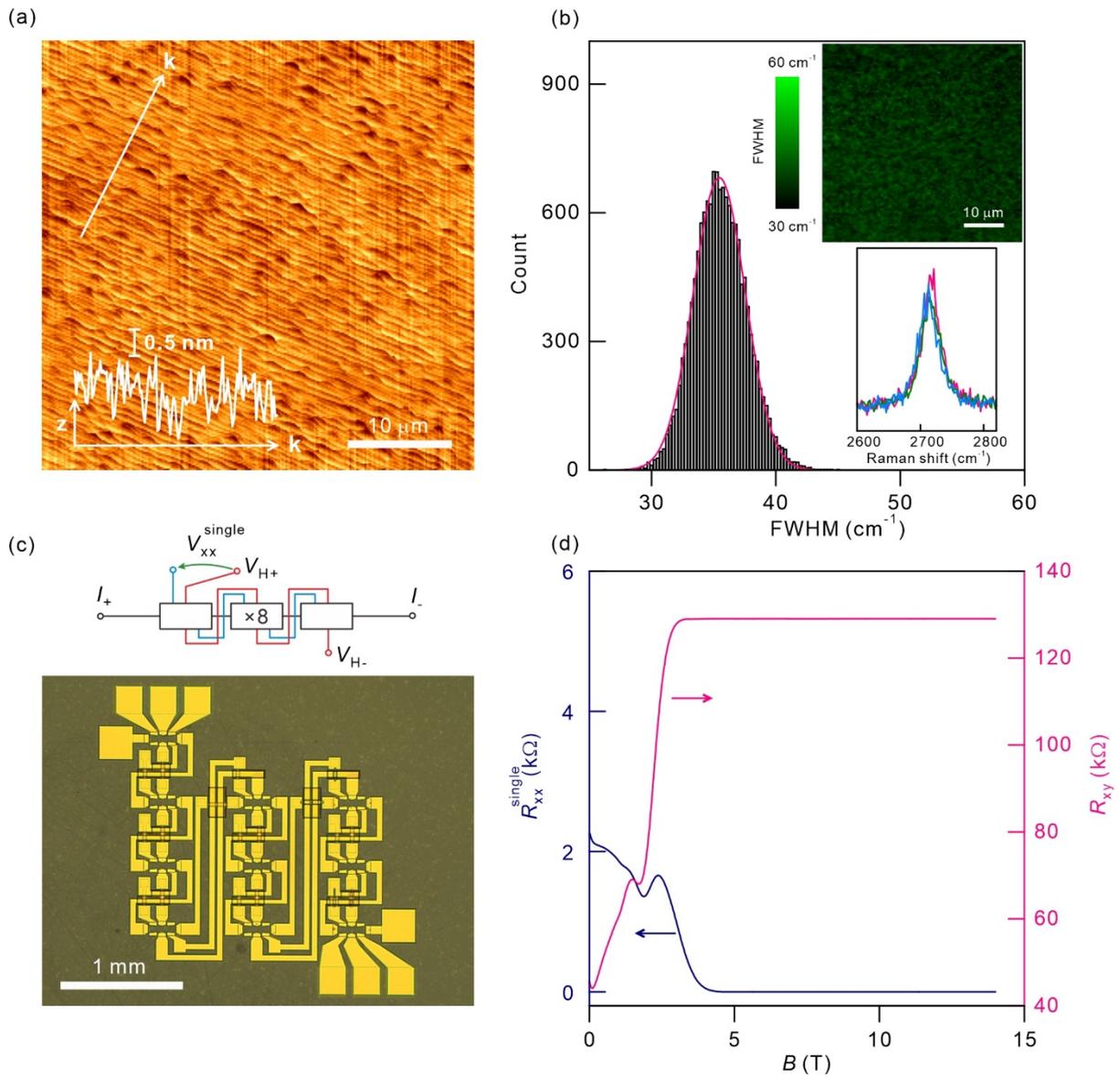

Fig. 1 (a) AFM image of EG on SiC. The inset shows the surface height along the white arrow line. (b) Count distribution of the 2D-FWHM. The upper and lower insets are Raman map of 2D-FWHM and Raman spectra of 2D-band on three points, respectively. Main histogram is extracted by counting the numbers of points in upper inset with 2D-FWHM. (c) Schematic diagram of 10 quantum Hall devices in series, linked by the triple connection geometry and the corresponding optical image of the device. (d) Magnetoresistance measurements acquired at 2 K.

**Figure 2**

(a)

(b)

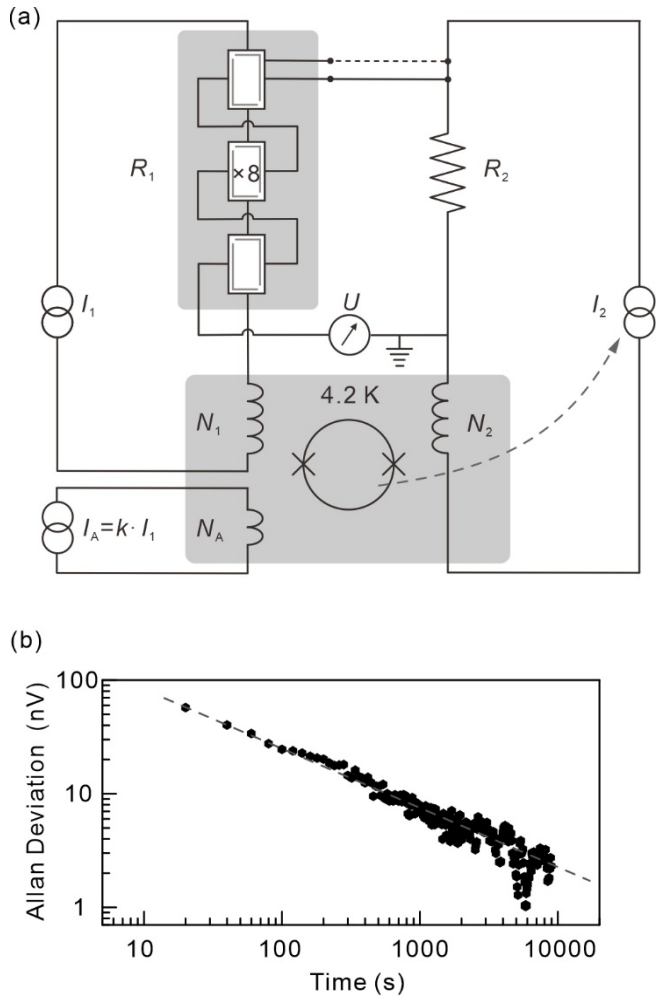

Fig. 2 (a) Schematic diagram of the cryogenic current comparator resistance bridge for precision measurements of $R_{xy}$ (configured by solid line) and $R_{xx}^{\text{single}}$ (configured by dotted line). Read details in main text. (b) Allan deviation of the bridge voltage difference, acquired for 10 hours at the magnetic field of 10 T and temperature of 2 K. It follows the inverse square root time dependence $(1/\sqrt{\tau})$ of white noise up to the sampling time of $10^4$ s.

**Figure 3**

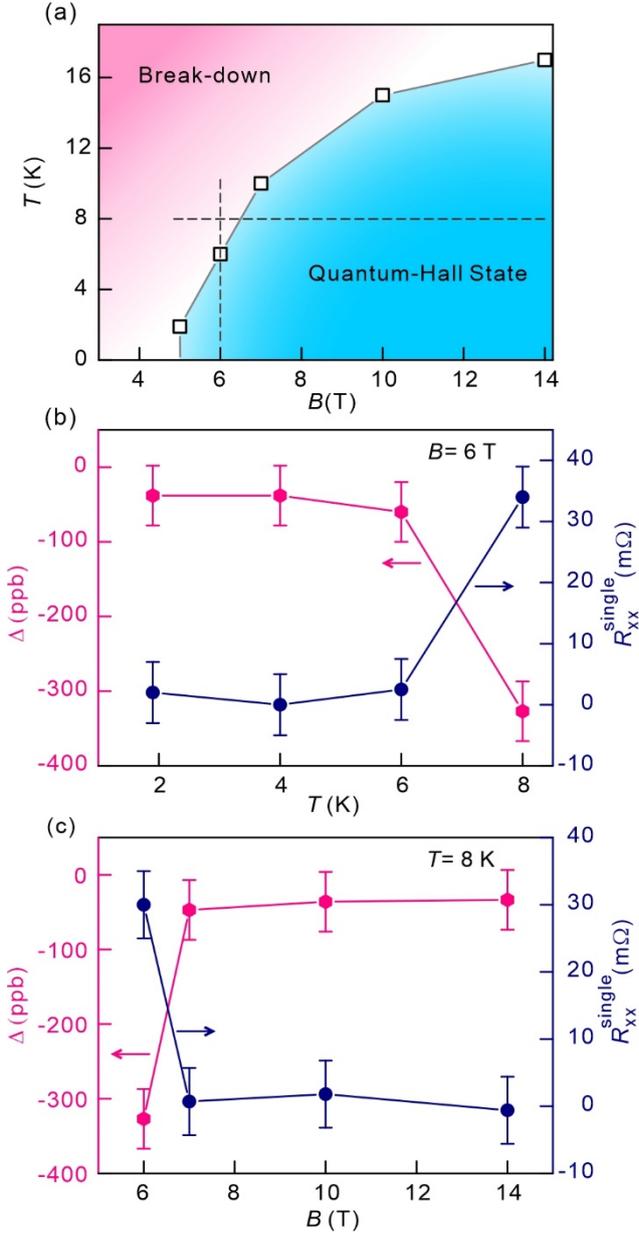

Fig.3 (a) Quantum-Hall phase diagram for graphene QHR array at filling factor 2 as a function of magnetic field and temperature. Within the blue region, the relative deviation of quantized Hall resistance in the array from $5\frac{h}{e^2}$ is smaller than $4\times10^{-8}$, comparable to the relative measurement uncertainty. (b) Red hexagonal(blue circular) symbol represents the relative deviation of $R_{xy}$ from $5\frac{h}{e^2}$ (the longitudinal resistance of single Hall element) with the magnetic field fixed at 6 T. $\Delta \equiv \frac{R_{xy}-5h/e^2}{5h/e^2}$. ppb stands for part per billion. (c) Red hexagonal(blue circular) symbol represents the relative deviation of $R_{xy}$ from $5\frac{h}{e^2}$ (the longitudinal resistance of a single Hall element) with the temperature fixed at 8 K.